\documentclass[twocolumn,twoside,slac_two]{revtex4}
\usepackage{graphicx}
\usepackage{fancyhdr}
\begin{document}

\title{Recent Lifetime and Mixing Measurements at the Tevatron}


\author{Chunlei Liu \\
(on behalf of CDF and D0 collaborations)}
\affiliation{University of Pittsburgh, Pittsburgh, PA }

\begin{abstract}
We present the latest $B$ hadron lifetimes, $B^0_s$ mixing, and $D^0$ mixing measurements
using up to 3.0~fb$^{-1}$ of data collected by CDF and D0 experiments at Fermilab. The $B^0_s$
lifetime is measured from both $J/\psi \phi$ ($CP$ admixture) and flavor specific channels, and
the $B_c$ lifetime is obtained from semileptonic channels. Following the $B^0_s$ oscillation 
frequency measurement at CDF in 2006, the D0 collaboration now observes $B^0_s$ oscillation at a significance
of about 3$\sigma$. Since the first $D^0$ mixing evidence established at $B$ factories in 2007, CDF
has observed $D^0$ mixing at a significance of about 4$\sigma$ level, the first time from hadron collider.   
\end{abstract}

\maketitle

\thispagestyle{fancy}


\section{Introduction}
Since the last FPCP conference in 2007, many interesting measurements have been carried at CDF and D0 experiments thanks to the
smoothly operating Tevatron. By March 30 2008, the Tevatron has delivered about 4~fb$^{-1}$ of integrated luminosity, while CDF 
and D0 recorded about 3.2 and 3.4 fb$^{-1}$ separately. The measurements related to these proceedings use data from 1.0~fb$^{-1}$ up to
2.8~fb$^{-1}$. The CDF and D0 experiments are described in detail in Ref.~\cite{ref:CDFd,ref:D0d}. For these proceedings, 
the recent $B^0_s$ and $B_c$ lifetimes results, $B^0_s$ and $D^0$ mixing parameters will be presented.  

\section{Lifetimes measurement}
\begin{figure}[htb]
\centering
\includegraphics[width=70mm]{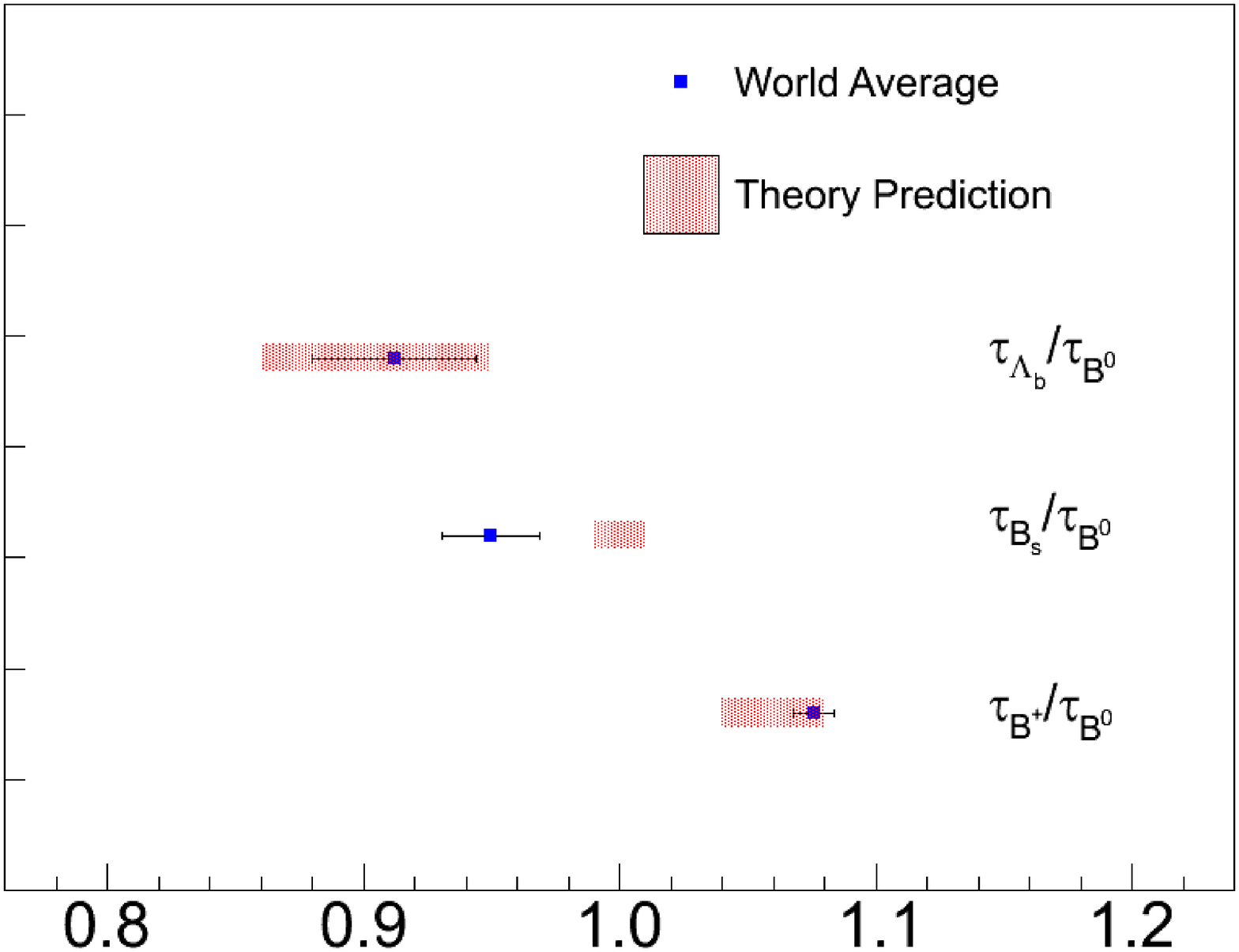}
\caption{Current lifetime ratios for different $B$ hadrons with theoretical predictions from Ref.~\cite{ref:timeratio}.} 
\label{fig:ratio}
\end{figure}

In the frame work of the Heavy Quark Expansion Theory (HQET), the inclusive decay rate of $B$ hadrons is given by~\cite{ref:HQE}:
\begin{equation}
\Gamma=\frac{G^2_F m^5_b}{192\pi^3}|V_{cb}|^2 \left[ \Gamma_0 +\Gamma_2 \left( \frac{\Lambda_{QCD}}{m_b} \right)^2+
                                                 \Gamma_3 \left( \frac{\Lambda_{QCD}}{m_b} \right)^3 \right ]
\label{eqn:lifetime}
\end{equation}
in the limit of $m_b\rightarrow \infty$ or a free quark model, all the $B$ hadrons should have same lifetime,
given by the first term of Eq.~\ref{eqn:lifetime}. The first correction arises from the kinetic and chromomagnetic
operator which is at order of $(\Lambda_{QCD}/m_b)^2$, the second correction arises from weak annihilation or Pauli
interference which is at order of $(\Lambda_{QCD}/m_b)^3$. Both theoretical and experimental uncertainties could be 
reduced if we measure the lifetime ratios of different $B$ hadrons.
Only the second correction will enter the lifetime ratio calculation. For example,
Pauli interference will increase $\tau(B^+)/\tau(B^0)$ and  $\tau(\Lambda_b)/\tau(B^0)$, and weak annihilation or scattering
will reduce it. As a result, the lifetime ratios are not exactly unity for different $B$ hadrons, and the precise 
measurements of the ratios can help test HQET at order of $(\Lambda_{QCD}/m_b)^3$. The current experimental situation
for $\tau_{B^+}/\tau_{B^0}$, $\tau_{B^0_s}/\tau_{B^0}$ and $\tau_{\Lambda_b}/\tau_{B^0}$ are shown in Fig.~\ref{fig:ratio},
where one can see that the experimental result and theoretical prediction agree well for the $B^+$ case. For the $B^0_s$
case, there is still some discrepancy between the prediction and measurements, while for the $\Lambda_b$, the theoretical prediction
range is large and the experimental results have large uncertainty at this time.

\subsection{$B^0_s$ lifetime measurements}
\begin{figure}[htb]
\centering
\includegraphics[width=60mm]{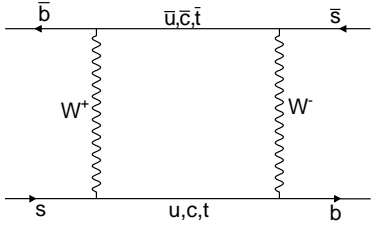}
\caption{Feynman diagram for $B^0_s-\bar{B}^0_s$ mixing. } 
\label{fig:box}
\end{figure}

The $B^0_s$ meson can decay directly or decay after it oscillates to $\bar{B}^0_s$. Time evolution of an arbitrary system 
$a(t)|B^0_s\rangle+b(t)|\bar{B}^0_s\rangle$ is governed by the Schr$\ddot{{\rm o}}$dinger Equation
\begin{equation}
i\frac{d}{dt} \left(\begin{array}{c}
a \\
b \end{array}\right) = \left(M-i\frac{\Gamma}{2}\right) \left(\begin{array}{c}
a \\
b \end{array}\right),
\label{eqn:Schrodinger}
\end{equation}
where $M$ and $\Gamma$ are mass and decay matrices. Two mass eigenstates $B^0_L$ and $B^0_H$
appear as a result of this mixing property. The eigenstates have different decay widths $\Gamma_L$ and $\Gamma_H$. The average 
decay width is defined as $\Gamma=(\Gamma_L+\Gamma_H)/2$, while the decay width difference is $\Delta\Gamma=\Gamma_L-\Gamma_H$.
$\Delta\Gamma$ also probes  new physics since it's related to quantity $\phi_s$ by $\Delta\Gamma=2|\Gamma_{12}|\cos(\phi_s)$, 
where $\phi_s={\rm arg}(-M_{12}/\Gamma_{12})$, while $\Gamma_{12}$ and $M_{12}$ are matrix elements coming from 
the box diagram as shown in Fig.~\ref{fig:box}. New physics is likely to increase $\phi_s$, so $\Delta\Gamma$ could be smaller than 
the Standard Model prediction.
 
\subsubsection{Results from $B^0_s\rightarrow J/\psi \phi$ channel}
\begin{figure}[htb]
\centering
\includegraphics[width=70mm]{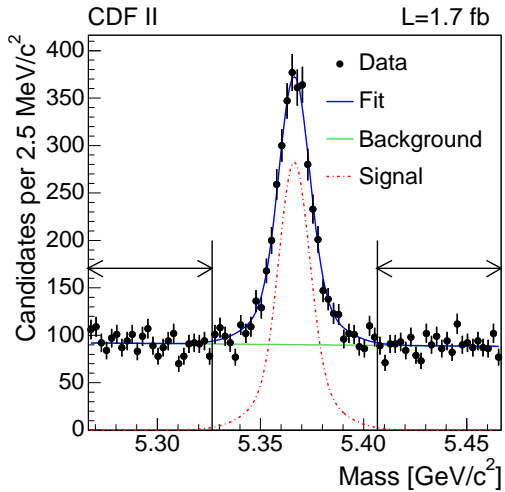}
\caption{The invariant mass distribution of $B^0_s$ candidates from CDF} 
\label{fig:Bs_mass_CDF}
\end{figure}

\begin{figure}[htb]
\centering
\includegraphics[width=70mm]{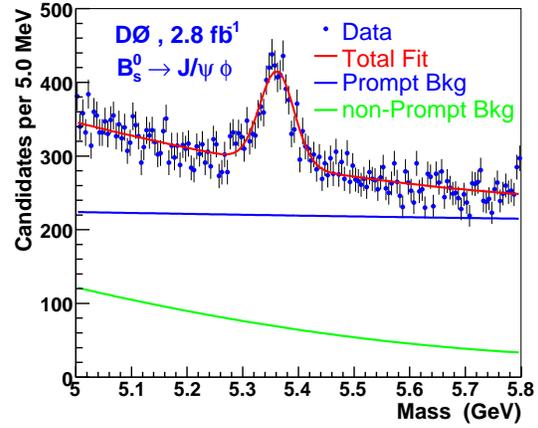}
\caption{The invariant mass distribution of $B^0_s$ candidates from D0 } 
\label{fig:Bs_mass_D0}
\end{figure}

The data at CDF (1.7~fb$^{-1}$) and at D0 (2.8~fb$^{-1}$) were collected by di-muon triggers~\cite{ref:CDFd,ref:D0d}  which preferentially 
selects events containing $J/\psi\rightarrow \mu^+ \mu^-$ decays. An artificial neural network is trained to separate signal and 
combinatorial backgrounds at CDF, which yields about 2500 signal events, while a cut based selection procedure is used to select data at D0,
giving about 2000 signal events. The mass projections from CDF and D0 are shown in Fig.~\ref{fig:Bs_mass_CDF} and 
Fig.~\ref{fig:Bs_mass_D0}.

Since $B^0_s$ is a pseudo-scalar meson, $J/\psi$ and $\phi$ are both vector mesons, the final states are admixtures of $CP$ 
eigenstates, where $S$ and $D$ waves are $CP$ even, $P$ wave is $CP$ odd. To separate the $CP$ eigenstates, the angular distribution 
are used~\cite{ref:AS}. $CP$ violation is predicted to be tiny in the Standard Model, so the $CP$ phase is fixed to zero at CDF,
while it's allowed to float at D0 with flavor tagging. To extract lifetime $\tau_s$ and decay width difference $\Delta\Gamma$, the final fit is 
done with an un-binned maximum likelihood method on mass, lifetime and  angular variables.

\begin{figure}[htb]
\centering
\includegraphics[width=70mm]{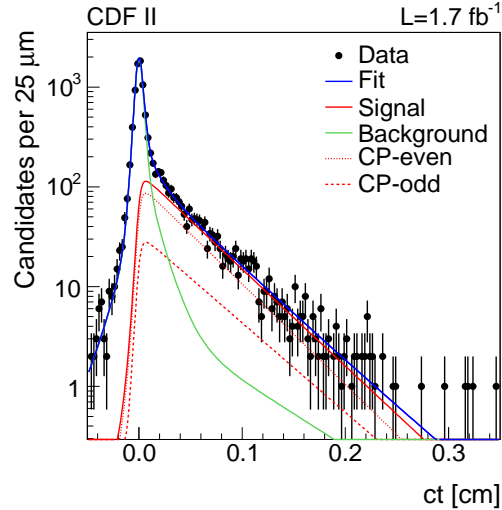}
\caption{Proper decay time projection of the $B^0_s\rightarrow J/\psi \phi$ fit at CDF. $CP$ even fraction is dominant, and 
the slope difference of even and odd curves indicates decay width difference.} 
\label{fig:Bs_life_CDF}
\end{figure}

\begin{figure}[htb]
\centering
\includegraphics[width=70mm]{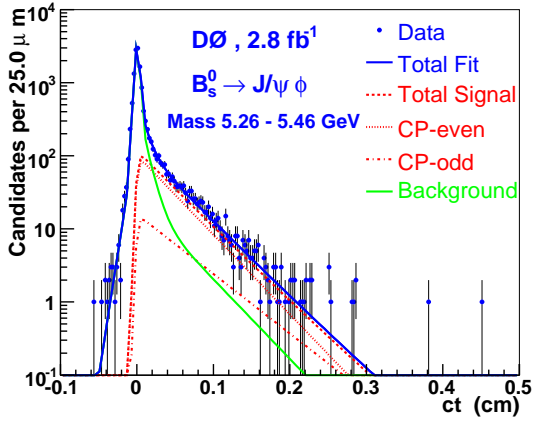}
\caption{Proper decay time projection of the $B^0_s\rightarrow J/\psi \phi$ fit at D0.} 
\label{fig:Bs_life_D0}
\end{figure}

The measured lifetime and decay width difference results from CDF are~\cite{ref:CDFbs}
\begin{eqnarray}
\tau= 1.52\pm0.04({\rm stat})\pm0.02({\rm syst})~ {\rm ps} \nonumber \\
\Delta\Gamma=0.08 \pm 0.06({\rm stat}) \pm 0.01({\rm syst})~ {\rm ps^{-1}} \nonumber
\end{eqnarray}
where $CP$ conservation is assumed, and the lifetime fit projection is shown in Fig.~\ref{fig:Bs_life_CDF}. 
At D0, the $CP$ violation phase is allowed to float (with external constraint on strong phases from $B^0\rightarrow J/\psi K^{*0}$), 
the results are~\cite{ref:D0bs}
\begin{eqnarray}
\tau= 1.52\pm0.05({\rm stat})\pm0.01({\rm syst}) ~{\rm ps} \nonumber \\
\Delta\Gamma=0.19 \pm 0.07({\rm stat})^{+0.02}_{-0.01}({\rm syst}) ~{\rm ps^{-1}} \nonumber
\end{eqnarray}
The lifetime fit projection is shown in  Fig.~\ref{fig:Bs_life_D0}.

The systematic errors are controlled very well at both CDF and D0. Systematic errors considered at CDF include background
angular distribution, mass model, lifetime resolution model, $B^0\rightarrow J/\psi K^{*0}$ contamination, detector acceptance, and
silicon detector alignment. Systematic errors considered at D0 include procedure test, acceptance, reconstruction algorithm, background
model, and detector alignment.

\subsubsection{Results from $B^0_s\rightarrow D_s(\phi \pi) \pi X$ channel}
\label{sec:b0s_semi}
This channel is different from $B^0_s\rightarrow J/\psi \phi$, because it's flavor specific, i.e. $B^0_s$ can only decay into 
$D^-_s \pi^+ X$, while 
$\bar{B}^0_s$ can only decay into $D^+_s \pi^- X$. By fitting the signal with one exponential function, the obtained lifetime is
related to average decay width and decay width difference by 
\begin{equation}
\tau(B^0_s)_{fs}= \frac{1}{\Gamma}(1+(\frac{\Delta\Gamma}{2\Gamma})^2)/(1-(\frac{\Delta\Gamma}{2\Gamma})^2)
\end{equation}
assuming no $CP$ violation. So it can be used to constrain $\Gamma$ and $\Delta\Gamma$ in the global fit. 

\begin{figure}[htb]
\centering
\includegraphics[width=70mm]{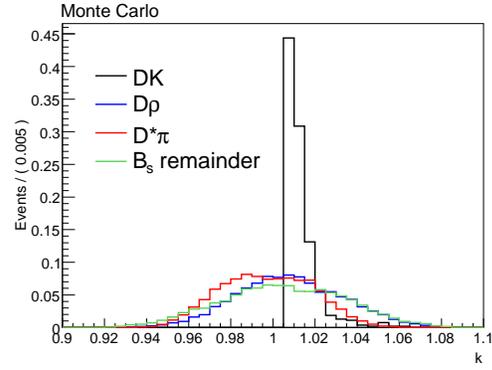}
\caption{K factor distribution for each channel. The $DK$ channel is fully reconstructed, but treated
in the same way, since the $K$ track is misidentified and a $K$ factor is needed.} 
\label{fig:bsKhist}
\end{figure}

The result is obtained at CDF only, from both fully reconstructed $B^0_s\rightarrow D_s \pi$ and  partially 
reconstructed channels such as  $B^0_s\rightarrow D_s \rho(\pi^+ \pi^0)$ where the $\pi^0$ is not reconstructed. In the
partially reconstructed channels, a ``K'' factor is introduced to correct the proper decay time for missing mass and  
transverse momentum
\begin{eqnarray}
ct= \frac{L_{xy}m^{rec}_B}{p_T} K \nonumber
\end{eqnarray}
where $L_{xy}$ is the projection of decay length in the $x-y$
plane, $m^{rec}_B$ is reconstructed mass, $p_T$ is reconstructed transverse momentum, and 
\begin{eqnarray}
K=\frac{m_Bp_T}{m^{rec}_B p^{true}_T} \nonumber
\end{eqnarray} 
where $m_B$ is the true mass and $p^{true}_T$ is the true transverse momentum. The ``K'' factor distributions for different 
partially reconstructed channels are obtained from Monte Carlo and  shown in Fig.~\ref{fig:bsKhist}.

The data (1.3~fb$^{-1}$) were collected by displaced-vertex trigger~\cite{ref:svt} at CDF. The trigger takes the advantage of 
the long lifetime of $B$ mesons and selects events within some impact parameter range ($120~\mu m\leq d_0 \leq 1000~\mu m$) 
with respect to the primary vertex.
The trigger makes it possible to select $B$ signal in large QCD backgrounds at CDF but also removes signal events which decay early.
The lifetime bias is corrected using a trigger efficiency curve obtained from Monte Carlo.

The lifetime is extracted from a two-step fit. First a mass-only fit is done to extract relative fractions of various modes and
backgrounds. The corresponding mass shapes are obtained from either Monte Carlo or data. A lifetime-only fit is performed with fractions 
fixed from the previous mass-only fit, where the lifetime is the only free parameter and the final result is~\cite{ref:bsfltime}
\begin{eqnarray}
\tau = 1.518 \pm 0.041({\rm stat}) \pm 0.025({\rm syst}) ~{\rm ps} \nonumber
\end{eqnarray}
The lifetime results for control samples, such as $B^0\rightarrow D^{-}(K^+\pi^-\pi^-)\pi^+$, 
$B^0\rightarrow D^{*-}(\bar{D}^0(K^+\pi^-)\pi^-)\pi^+$, and $B^+\rightarrow \bar{D}^0(K^+\pi^-)\pi^+$, all agree
well with PDG values. 

\begin{figure}[htb]
\centering
\includegraphics[width=70mm]{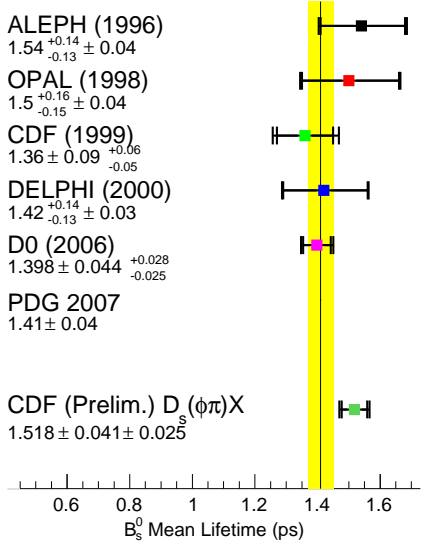}
\caption{$B^0_s$ lifetime results  from flavor specific channels. }
\label{fig:bssemi}
\end{figure}

Fig.~\ref{fig:bssemi} compares this result with all published results from flavor specific channels. The previous PDG 
value is dominated by the D0 result in 2006 from $B^0_s$ semileptonic channels. This new result agrees with the PDG 2007 value,
but the central value is higher.

\subsection{$B_c$ lifetime measurement}
The doubly heavy $B_c$ meson is an interesting QCD laboratory, where both $b$ and $c$ quarks can decay as well as annihilate.
The lifetime is expected to be much shorter than light $B$ mesons. Current theory predictions span from 0.47~ps 
to 0.59~ps~\cite{ref:bctheory}, so experimental result with small uncertainty will be useful to constrain the theories.

Both CDF and D0 have measured the $B_c$ lifetime from $B_c\rightarrow J/\psi l X$ semileptonic channels, and data are collected by the
di-muon trigger. At CDF both $B_c\rightarrow J/\psi \mu X $ and  $B_c\rightarrow J/\psi e X $ decays are reconstructed from 1.0~fb$^{-1}$
data. At D0 only $B_c\rightarrow J/\psi \mu X $ is reconstructed from 1.3~fb$^{-1}$ data. Because of the missing momentum of the neutrino,
a ``K'' factor is also needed from Monte Carlo to correct the reconstructed lifetime value.    
Here the mass is taken from the measured value~\cite{ref:bcmass} instead of 
the reconstructed value, so 
the ``K'' factor takes the form
\begin{eqnarray}
K=\frac{p_T(J/\psi l) }{p_T(B_c)} \nonumber
\end{eqnarray}

The main challenge of the analysis is the various and dominant backgrounds which include: 1) fake $J/\psi$ plus true lepton, 2) true
$J/\psi$ plus fake lepton, 3) true $J/\psi$ and lepton from  different $b$ quarks, 4) prompt $J/\psi$ background, 5) residual 
conversion (for electron channel only). At CDF, the shape of the proper decay time distributions of all backgrounds are modeled and 
calibrated carefully from either data or Monte Carlo, and the lifetime is extracted from a lifetime-only fit. At D0, 
the mass shape of signal and backgrounds are also modeled, and the lifetime is extracted from a mass-lifetime simultaneous fit. 

\begin{figure}[htb]
\centering
\includegraphics[width=70mm]{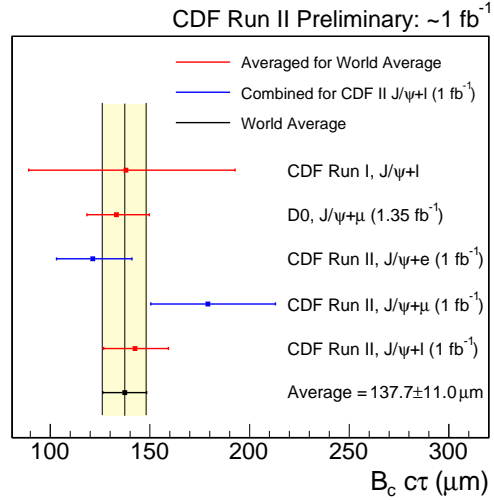}
\caption{Comparison of $B_c$ lifetime with CDF and D0 RunI results, and an weighted average (assuming no correlations 
among) those measurements.}
\label{fig:bcworld}
\end{figure}

The measured lifetime results at CDF from both muon and electron channels are~\cite{ref:CDFbc}
\begin{eqnarray}
c\tau = 179.1^{+32.6}_{-27.2}({\rm stat})~ \mu m  \     ({\rm muon\ channel}) \nonumber \\
c\tau = 121.7^{+18.0}_{-16.3}({\rm stat})~ \mu m  \    ({\rm electron\ channel}) \nonumber
\end{eqnarray}
and the combined result from two channels is given by
\begin{eqnarray}
\tau(B_c)= 0.475^{+0.053}_{-0.049}({\rm stat})\pm0.018({\rm syst})  ~{\rm ps}    \nonumber
\end{eqnarray}
The same lifetime measured at D0 is~\cite{ref:D0bc}
\begin{eqnarray}
\tau(B_c)= 0.448^{+0.038}_{-0.036}({\rm stat})\pm0.032({\rm syst})  ~{\rm ps}     \nonumber
\end{eqnarray}
The measured $B_c$ lifetime values from CDF and D0 agree with each other well, and since the Tevatron is currently the only place
which can produce $B_c$ mesons, one can make a weighted world average with all the previous measurements from the Tevatron.
The plot is shown in Fig.~\ref{fig:bcworld}, and the weighted average is
\begin{eqnarray}
\tau(B_c)= 0.459\pm0.037 ~{\rm ps} \    \nonumber
\end{eqnarray}

\section{$B^0_s$ Mixing}

As shown in Fig.~\ref{fig:box}, $B^0_s$ oscillates through a box diagram. The oscillation frequency $\Delta m_s$ is the mass difference
of heavy and light eigenstates $\Delta m_s\equiv m_H-m_L$, which are related to the off diagonal element $M_{12}$ of the mass matrix by
\begin{equation}
\Delta m_s= 2|M_{12}|
\end{equation}
With the oscillation frequency from the $B^0$ system, one can take the ratio
to cancel most of the theoretical uncertainties and get
\begin{equation}
\frac{\Delta m_s}{\Delta m_d} \propto \frac{|V_{ts}|^2}{|V_{td}|^2} 
\end{equation} 
which is very important for constraining the CKM unitary triangle with measurements of other CKM matrix elements and some theory inputs. 

The probability for a $B^0_s$ meson to mix or not as a function of time is proportional to
\begin{equation}
P(t)_{B^0_s\rightarrow B^0_s, \bar{B}^0_s} \propto 1\pm \cos{\Delta m_s t}
\label{eqn:oscillation}
\end{equation}
And one needs to identify the flavor of the meson at both production and decay time to see if it's mixed or unmixed.
By reconstructing flavor specific channels, the flavor at decay time can be identical to the charge of 
the daughter particle. To identify the flavor at production time, two algorithms are generally used at the Tevatron. On the same side
of the reconstructed $B^0_s$ meson, one looks at the charge of the kaon  from fragmentation processes.
On the opposite side, one can look at the charge of leptons from semileptonic decays or the jet charge
of the other $b$ quark. However, many things can dilute the tagging power, for example, the kaon mis-identification and low efficiency 
on the same side, mixing of neutral $B$ mesons and sequential $b$ decays on the opposite side. Thus, a dilution factor ${\cal D}$ is 
introduced, and it modifies Eq.~\ref{eqn:oscillation}.
\begin{equation}
P(t)_{B^0_s\rightarrow B^0_s, \bar{B}^0_s} \propto 1\pm {\cal D}\cos{\Delta m_s t}
\end{equation}  
and $(1+{\cal D})/2$ gives correct tagging probability.

Following the CDF measurement in 2006 with $\Delta m_s = 17.77 \pm 0.10 ({\rm stat}) \pm 0.07 ({\rm syst})~{\rm ps}$, D0 now has a new measurement 
with 2.4~fb$^{-1}$ data. The first 1.3~fb$^{-1}$ belongs to Run IIa period and the other 1.1~fb$^{-1}$ belongs to Run IIb when an additional
silicon layer was installed at D0. Both hadronic and semileptonic decays are collected by inclusive single or di-muon triggers at D0. 
Most of the reconstructed channels and signal yields are listed in Tab.~\ref{table:bs_yield}, in addition, 1.2~fb$^{-1}$ of 
$B^0_s\rightarrow \mu^+ D^-_s(K^0_s K^-)$ decays from Run IIa give about 600 signal events.  

\begin{table}[h]
\begin{center}
\caption{Reconstructed channels and yields in $B^0_s$ mixing measurement}
\begin{tabular}{|l|c|c|c|}
\hline
 \textbf{Channel} & \textbf{Run IIa} & \textbf{Run IIb} & \textbf{Total}
\\
\hline 
 $\mu^+ D^-_s(\phi \pi^-)$ &  $28238\pm339$  & $16539\pm239$  & $44777\pm415$ \\
\hline 
 $e^+ D^-_s(\phi \pi^-)$ &  $1142\pm83$  & $548\pm45$  & $1663\pm102$ \\
\hline
 $\pi^+ D^-_s(\phi \pi^-)$ &  $159\pm13$  & $90\pm11$  & $249\pm17$ \\
\hline 
 $\mu^+ D^-_s(K^{*0} K^-)$ &  $11649\pm661$  & $6449\pm616$  & $18098\pm903$ \\ 
\hline
\end{tabular}
\label{table:bs_yield}
\end{center}
\end{table}

\begin{figure}[htb]
\centering
\includegraphics[width=70mm]{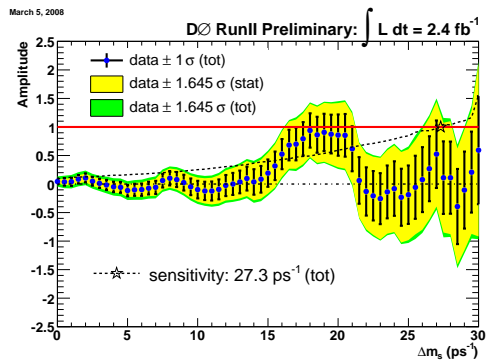}
\caption{$B^0_s$ mixing combined amplitude scan.} \label{fig:bs_amp}
\end{figure}

\begin{figure}[htb]
\centering
\includegraphics[width=70mm]{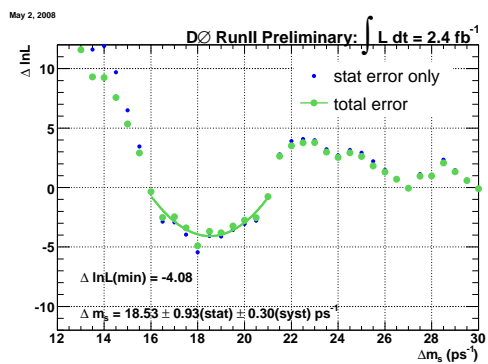}
\caption{$B^0_s$ combined likelihood profile, which is done with statistical error and total error on amplitude $A$ 
separately. An estimate of the $\Delta m_s$ systematic error is obtained by subtracting, in quadrature, the error
derived from statistical-error-only likelihood curve from that of the total-error curve. } 
\label{fig:bs_ll}
\end{figure}

An amplitude $A$ is introduced as shown in 
\begin{equation}
P(t)_{B^0_s\rightarrow B^0_s, \bar{B}^0_s} \propto 1\pm A{\cal D}\cos{\Delta m_s t}
\end{equation} 
where $A$ can be fitted by fixing $\Delta m_s$ at different points, the true $\Delta m_s$ value is indicated  when the fitted A 
is consistent with unity, otherwise  a sensitivity at 95\% C.L. can be defined by the probe $\Delta m_s$ value at which
\begin{equation}
1.645 \sigma_A(\Delta m_{sens}) =1.0
\end{equation}
Individual amplitude scans vs. probe $\Delta m_s$ were done separately for different channels and combined using 
COMBOS program~\cite{ref:combine}. The combined amplitude scan is shown in Fig.~\ref{fig:bs_amp}. The likelihood profile
is obtained from the combined amplitude scan using the formula~\cite{ref:likelihood}
\begin{equation}
\Delta Ln {\cal L} = - Ln {\cal L}(\Delta m_s) + Ln {\cal L}(\infty)=\frac{\frac{1}{2}-A_{fit}(\Delta m_s)}{\sigma^2_A}
\end{equation}
and the profile is shown in Fig.~\ref{fig:bs_ll}. The measured $\Delta m_s$ and its errors are derived by fitting a quadratic 
function to it, giving~\cite{ref:bsmixing}
\begin{equation}
\Delta m_s = 18.53 \pm 0.93({\rm stat}) \pm 0.30({\rm syst}) 
\end{equation}
which has a total significance of 2.9~$\sigma$.

\section{$D^0$ Mixing}

Since the discovery of charm quark in 1974, physicists have been trying to observe oscillations
of the neutral charm meson. Unlike the kaon or $B$ systems, the 
oscillation frequency in the charm system is predicted to be very small. In the Standard Model, $D^0$ mixing occurs 
through two processes. The ``long range'' mixing can be estimated using 
strong interaction models. 
The ``short range'' process is negligible in the standard model, however, exotic weakly interacting particle from
new physics could enhance the short-range process. 

Recent $D^0$ mixing evidence has been found at the $B$ factories in 2007. The $Belle$ Collaboration found direct
evidence by comparing the decay time distributions for $D^0$ decays to the $CP$-eigenstates $K^+K^-$ and $\pi^+\pi^-$ 
with that for the $CP$-mixed state $K^-\pi^+$. The evidence found in the $BABAR$ experiment is the decay rate difference
of doubly-Cabibbo-suppressed (DCS) $D^0\rightarrow K^+\pi^-$ decay and Cabibbo favored (CF) $D^0\rightarrow K^-\pi^+$ 
decay. The ratio of decay rates can be approximated as a simple quadratic function of proper decay time and mean 
$D^0$ lifetime, with the assumption of $CP$ conservation and a small value of $x=\Delta m/\Gamma$ and $y=\Delta\Gamma/2\Gamma$,
\begin{equation}
R(t/\tau)=R_D + \sqrt{R_D}y^{\prime}(t/\tau)+\frac{x^{\prime2}+y^{\prime2}}{4}(t/\tau)^4
\label{eqn:d0mixratio}
\end{equation}
where $R_D$ is the square of the ratio of DCS to CF amplitudes, $x^{\prime}$ and $y^{\prime}$ are linear combinations of 
$x$ and $y$ as
\begin{eqnarray}
x^{\prime}=x\cos\delta+y\sin\delta \nonumber \\
y^{\prime}=-x\sin\delta+y\cos\delta \nonumber 
\end{eqnarray}
where $\delta$ is the strong phase difference between the DCS and CF amplitudes. In the absence of mixing, $x^{\prime}$
and $y^{\prime}$ are both zero.

\begin{figure}[htb]
\centering
\includegraphics[width=70mm]{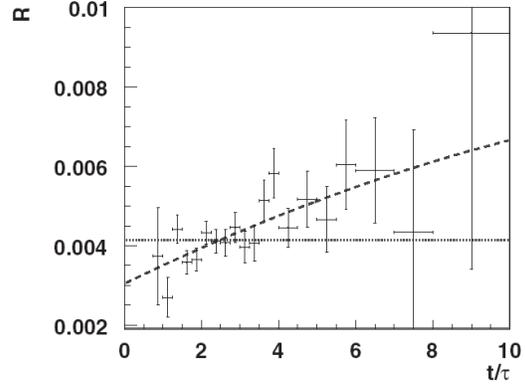}
\caption{Ratio of prompt $D^*$ ``wrong sign'' to ``right sign''decays as a function of normalized proper decay time.
The dashed curve is from a least-squares parabolic fit, the dotted line is the fit assuming no mixing } 
\label{fig:d0time}
\end{figure}

At CDF, the same kind of evidence is found as in $BABAR$, but probed over a much wider $D^0$ decay time range.
Events are selected with the displaced-vertex trigger using about 1.5~fb$^{-1}$ data. The ``right-sign'' (RS) CF decay chain
$D^{*+}\rightarrow \pi^+ D^0$, $D^0\rightarrow K^-\pi^+$, and the ``wrong sign'' (WS) decay chain $D^{*+}\rightarrow \pi^+ D^0$,
$D^0\rightarrow K^+\pi^-$ are reconstructed. The relative charges of the two pions gives the ``sign'' of the decay chain and 
extra cuts are applied to reduce background to the WS signal from RS decays. The data are divided into 20 bins of $t/\tau$ ranging
from 0.75 to 10.0 with bin size increasing from 0.25 to 2.0 to reduce statistical uncertainty at larger times. The ratio $R$
is determined at each bin and a least-squares parabolic fit of the data is done with Eq.~\ref{eqn:d0mixratio}. The fit is shown
in Fig.~\ref{fig:d0time}, and final results are shown in Table.~\ref{table:d0fit}~\cite{ref:d0mixing}.

\begin{table}[h]
\begin{center}
\caption{Fit results for the $R(t/\tau)$ distribution}
\resizebox{8cm}{!}{
\begin{tabular}{|l|c|c|c|c|}
\hline
 \textbf{Fit type} & \textbf{$R_D(10^{-3})$} & \textbf{$y^{\prime}(10^{-3})$} & \textbf{$x^{\prime 2}(10^{-3})$} 
 & \textbf{$\chi^2/d.o.f$}
\\
\hline 
 Unconstrained &  $3.04\pm0.55$  & $8.5\pm7.6$  & $-0.12\pm0.35$  & 19.2/17\\
\hline 
 Physically allowed &  $3.22\pm0.23$  & $6.0\pm1.4$  & $0$  & 19.3/18\\
\hline
 No mixing &  $4.15\pm0.10$  & $0$  & $0$  & 36.8/19\\
\hline 
\end{tabular}
}
\label{table:d0fit}
\end{center}
\end{table}

\begin{figure}[]
\centering
\includegraphics[width=70mm]{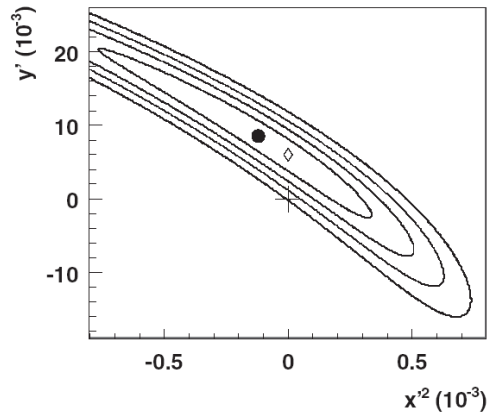}
\caption{Bayesian probability contours in the $x^{\prime 2}-y^{\prime}$ parameter space corresponding to one
through four equivalent Gaussian standard deviations. The closed circle shows the unconstrained fit, the open diamond
shows physically allowed fit ($x^{\prime 2}\geq 0$), the cross shows the no-mixing point. } 
\label{fig:d0contour}
\end{figure}

To determine the consistency of the data with the no-mixing hypothesis, Bayesian contours are computed with
likelihood and a flat prior. The contours are insensitive to modest changes in the prior, and  shown in 
Fig.~\ref{fig:d0contour}. The no-mixing point lies on the contour which excludes the region containing a probability 
of $1.5\times10^{-4}$, equivalent to 3.8 Gaussian standard deviations. Alternative procedures are also used to 
determine the probability for no mixing, and  all give consistent results.
 
\section{Summary}
The most precise $B^0_s$ lifetime and decay width difference have been measured directly from the $B^0_s\rightarrow J/\psi \phi$ channel
from both CDF and D0 experiments. The measured lifetime is also consistent with the value from $B^0_s$  flavor specific channel.
Together, the recent results confirm the theory prediction that $\tau_s/\tau_d\sim 1$. The lifetime measurement result for 
the heavy $B_c$ meson can provide good experimental input for the theory calculation which still has a huge uncertainty. The fast 
oscillation of $B^0_s$ mesons has been measured at CDF in 2006 and is now verified by the D0 experiment with results from combined
channels. While $D^0$ mixing is predicted to be quite small in the Standard Model, the evidence has been found at CDF followed the 
first observation at $B$ factories last year. 
 
\begin{acknowledgments}
I would like to thank the organizers of FPCP 2008 conference for a enjoyable week. I also would like to thank all the colleagues at 
both CDF and D0 collaborations and Fermilab staff for their hard work which makes these results possible. 
\end{acknowledgments}

\bigskip % extra skip inserted

\end{document}